\documentclass[twocolumn,showpacs,aps]{revtex4}
\usepackage[dvips]{graphicx}
\usepackage{bm}
\usepackage{mathrsfs}

\begin{document}
\date{\today}

\title{From Fermat Principle to Wave equation\\
-Quantization of ``particle mechanics of Light"-}
\author{Naohisa Ogawa
\footnote{ogawanao@hit.ac.jp}}
\affiliation{Hokkaido Institute of Technology, Sapporo 006-8585 Japan}
\begin{abstract}
The Fermat principle indicates that light chooses the temporally shortest path. 
The action for this ``motion" is the observed time, and it has no Lorentz invariance. 
In this paper we show how this action can be obtained from relativistic action, 
and how the classical wave equation of light can be obtained from this action.
\end{abstract}
\pacs{03.65.Pm}
\maketitle

\section{Massless limit of relativistic particle and dimensional reduction}

Let us start with the action of relativistic massive particle, i.e. the proper time
\begin{equation}
S_0 = -m \int d\tau \sqrt{\dot{x}^\mu \dot{x}_\mu}.
\end{equation}
But we can not take massless limit of this action.
So we use the different action which is equivalent to $S_0$
\begin{equation}
S_1 = -\frac{1}{2} \int d\tau \sqrt{g} (g^{-1} \dot{x}^\mu \dot{x}_\mu + m^2),
\end{equation}
where $g$ is the auxiliary field.
Let us show the equivalence of these two actions.
The Euler-Lagrange equation for $g$ is
\begin{equation}
g=\frac{\dot{x}^\mu \dot{x}_\mu}{m^2}.
\end{equation}
By putting above equation for $g$ into the action $S_1$, we obtain $S_0$.
So we observe the equivalence.
To  consider massless limit, we use $S_1$ instead of $S_0$.
The action for massless particle is then,
\begin{equation}
S_2 = -\frac{1}{2} \int d\tau g^{-1/2} \dot{x}^\mu \dot{x}_\mu.
\end{equation}
Two Euler-Lagrange equations are
\begin{equation}
\frac{d}{d\tau}(\frac{\dot{x}^\mu}{\sqrt{g}})=0, ~~~~~\dot{x}^\mu \dot{x}_\mu=0. \label{eq:EL}
\end{equation}
The latter equation gives
\begin{equation}
\dot{x}^0 = \pm \sqrt{\dot{\bf{x}}^2}.
\end{equation}
Then the first equation of (\ref{eq:EL}) becomes
\begin{equation}
\frac{d}{d\tau}(\frac{\sqrt{\dot{\bf{x}}^2}}{\sqrt{g}})=0,~~~~~\frac{d}{d\tau}(\frac{\dot{x}^i}{\sqrt{g}})=0.\label{eq:EL2}
\end{equation}
The first equation shows that ratio of $g$ and $\dot{\bf{x}}^2$ is constant. 
And so by using constant $\lambda$, we can write
\begin{equation}
g= \lambda \dot{\bf{x}}^2.
\end{equation}
By putting this into the second equation of (\ref{eq:EL2}), we have
\begin{equation}
\frac{d}{d\tau}(\frac{\dot{x}^i}{\sqrt{\dot{\bf{x}}^2}})=0. \label{eq:EL3}
\end{equation}
The action which gives (\ref{eq:EL3}) as Euler-Lagrange equation is
\begin{equation}
S_3 = \frac{1}{c} \int d\tau \sqrt{\dot{\bf{x}}^2},
\end{equation}
where the speed of light $c$ is introduced for the action to have the dimension of time.
$S_3$ is the time along the path of light, and it is really the action expressing the Fermat principle.
Though it seems non-relativistic, it should introduce the wave equation of light.
This point will be made clear in the following sections.
Note that the dimension of configuration space in $S_3$ is 1 dimension less than the one in $S_0$.
The massless limit induces the dimensional reduction in action, which is similar to the change of  physical degree of freedom: 
from massive vector field (3 degrees of freedom) to massless vector field (2 degrees of freedom).
Another point is that action $S_3$ still have time-reparametrization invariance \cite{Polyakov}\cite{others}\cite{ishikawa}.
Due to this local symmetry, the method to obtain the wave equation becomes bit complicated.
This is  discussed in the next section.

\section{Time reparametrization symmetry}

Let us consider the problem to obtain the wave equation of light from action $S_3$.
But $S_3$ is not enough to express the Fermat principle.
Because the minimisation of $S_3$ can also be understood as minimisation of path-length.
To make clear the meaning of minimisation of time along the path, we generalize the action with refractive index field as
\begin{equation}
  S \propto ``time" = \frac{1}{c} \int^{T}_{0} 
  n(x)[(\frac{d {\bf x}}{d \tau})^2]^{1/2} d \tau \label{eq:action},
\end{equation}
where $c$ is the speed of light, $n(x)$ is the refractive index and $\tau$ is only the parameter but not time.
$d\tau [(\frac{d {\bf x}}{d \tau})^2]^{1/2}$ is the infinitesimal spatial length along the light path. 
This action is invariant under the transformation $$\tau \to f(\tau),$$
if the conditions $df/d \tau > 0,~~0=f(0)$, and $T=f(T)$ hold.
This is well known as time reparametrization \cite{Polyakov},\cite{others}.
The equation of motion as ``particle mechanics of light" is
\begin{equation}
\frac{d}{d \tau}[n(x) \frac{\dot{\bf x}}{\sqrt{ \dot{\bf x}^2}}] 
= \sqrt{ \dot{\bf x}^2} ~ \nabla n(x). \label{eq:motion}
\end{equation}
This equation can be re-expressed by using the length parameter $s$; $ds=\sqrt{\dot{\bf x}^2}d\tau$,
\begin{equation}
\frac{d^2 \vec{x}}{ds^2} = \frac{1}{n}[ \vec{\nabla}n - (\vec{\nabla} n \cdot \frac{d\vec{x}}{ds}) \; \frac{d\vec{x}}{ds}].
\end{equation}
 
The right hand side can be interpreted as some-kind of ``force".
The property of this force is the followings.
When $\vec{\nabla} n // (d\vec{x}/ds)$, then $\vec{\nabla} n = \mid \vec{\nabla} n \mid d\vec{x}/ds$, 
and the force vanishes.
When $\vec{\nabla} n \perp d\vec{x}/ds$, then r.h.s. is $\vec{\nabla} n/n$.
Further, (\ref{eq:action}) can be expressed as
\begin{equation}
S \propto \int^{T}_{0} \sqrt{g_{ij} \dot{x}^i \dot{x}^j} d\tau,
\end{equation}
with conformally flat metric: $g_{ij}=(n/c)^2 \delta_{ij}.$
Then the equation of geodesic line is the following.
\begin{equation}
\frac{d^2 x^i}{d\bar{s}^2} + \Gamma^i_{jk} \frac{dx^j}{d\bar{s}} \frac{dx^k}{d\bar{s}}=0,
\end{equation}
where $d\bar{s}=\frac{n(x)}{c}ds$. This equation is equivalent to (\ref{eq:motion}), where the Christoffel Symbol is
$$ \Gamma^i_{jk} = \frac{1}{n}(\delta^i_j \frac{\partial n}{\partial x^k} + \delta^i_k \frac{\partial n}{\partial x^j} - 
\delta_{jk} \frac{\partial n}{\partial x^i}).$$

Our program is the following.
To consider the quantization procedure, it is necessary to construct dimensionless action.
This is done by multiplying the frequency of light: $\omega$ (monochrome light) .
We have
\begin{equation}
  S = \frac{\omega}{c} \, \int^{T}_{0} \,  \;n(x) 
[(\frac{d {\bf x}}{d \tau})^2]^{1/2} d \tau .\label{eq:action2}
\end{equation}

Then we consider the path-integration,\cite{Wentzel},\cite{Feynman}
\begin{equation}
\Psi[x_f, x_i, \omega]_{st} = \sum_{path C} e^{iS[C]}.\label{eq:path}
\end{equation}

To obtain the time which we observe, we sum up the function $\Psi$ with various frequencies.
\begin{equation}
\Psi[x_f,t : x_i,0] = \frac{1}{2\pi}\int d\omega  \Psi[x_f, x_i, \omega]_{st} e^{-i\omega t}.
\end{equation}
Then we seek the wave equation which $\Psi[x_f,t : x_i,0] $ obeys.
Note that we have no $\hbar$ in the path-integration. 
This is because we are treating Light: classical wave instead of Schr\"odinger equation.

The explicit form of path-integral (\ref{eq:path}) is the following.
\begin{equation}
\Psi_{st}= \int {\cal D}x \exp[i\frac{\omega}{c} \int^{T}_{0} n(x)
\{(\frac{d {\bf x}}{d \tau})^2\}^{1/2}d\tau ].
\end{equation}
But the treatment of this style action is not convenient. 
So we change the action into the canonical form.
From the original action~$S$, we can define the canonical momentum $ p_{i}(\tau)$ 
which is conjugate to $x^{i}$. Then we have the Hamiltonian constraint 
${\cal{H}} \equiv p^2 - (n\omega/c)^2 $, and the canonical action.

\begin{equation}
S  = \int^{T}_{0} \, d \tau \, L(\tau) \: = \: 
\int^{T}_{0} \, d \tau \; [\: p_{i}\, 
\dot{x}^{i} \, - \, N(p^2 - \omega^2 n^2/c^2)\:],
\end{equation}
where $N$ is the Lagrange multiplier field.
The equation of motion from this action is the same as the previous one (\ref{eq:motion}).
Time reparametrization invariance exists also in this action.
The transformation property for each field is
\begin{eqnarray}
   x^{i}(\tau)  & \longrightarrow & x^{i}(f(\tau)),\\
   p_{i}(\tau)  & \longrightarrow & p_{i}(f(\tau)),\\
   N(\tau)        & \longrightarrow & N(f(\tau))\, \dot{f}(\tau).
\end{eqnarray}
This symmetry  is  canonically induced from the first class constraint ${\cal H}$. \cite{others}
The canonical infinitesimal transformation is given as
\begin{eqnarray}
\delta x^{i}\: & \equiv & \: \{ \:x^{i},\: p^2-\omega^2 n^2/c^2\: \} \, \varepsilon \:
 = \: 2 \,p_{i}\, \varepsilon ,\\ 
\delta p_{i}\: & \equiv & \: \{ \:p_{i},\: p^2-\omega^2 n^2/c^2\: \} \, \varepsilon \:
 = \: 2\omega^2 n \partial_i n \, \varepsilon /c^2 .
\end{eqnarray}
The transformation for $N(\tau)$ is defined for action to be invariant as
\begin{equation}
 \delta N \: \equiv \: \dot{\varepsilon}.\label{eq:rep}
\end{equation}

The symmetry is as follows:
\begin{equation}
 \delta L(x,p,N) = \frac{d}{d \tau}\:[\,\varepsilon(\tau)\:(\, p^2 \,+ \, \omega^2 n^2/c^2 \,)\,].
\end{equation}
The invariance of action requires $\varepsilon(0)=\varepsilon(T)=0$, 
as the boundary condition which we will use later. 

\section{Path-Integration}
Under the transformation (\ref{eq:rep}), the naive measure for $N$ is invariant.
So we can safely consider the path-integral:
\begin{equation}
 \Psi_{st} \: = \: \int {\cal D}x \:{\cal D}p \: {\cal D}N e^{iS}.
\end{equation}
Let us use the gauge condition: $\dot{N}=0$ that fix the 
functional form of infinitesimal parameter as
\begin{equation}
               \dot{N} = 0  \rightarrow  \ddot{\varepsilon} = 0,
\end{equation}
which is the similar relation as Lorentz gauge in U(1) gauge theory.
Then we can construct the Faddeev-Popov determinant as:
\begin{equation}
\Delta^{-1}[N] \; \equiv \; \int {\cal D}\varepsilon \; \delta (\dot{N}^{\varepsilon}),
\end{equation}
where $N^\varepsilon$ means the gauge transformed field $ N(\tau)$ by time-reparametrization
 $\varepsilon$, and ${\cal D}\varepsilon$ is the Haar measure of time-reparametrization group.
Then our Path-integration has the form
\begin{equation}
   \Psi_{st} = \int  {\cal D}x \, {\cal D}p \, {\cal D}N \,  \Delta[N] \: 
\delta(\dot{N}) e^{iS}.
\end{equation}
$\Delta[N]$ can be calculated under the gauge condition $\dot{N} = 0$ with 
infinitesimal parameter $\varepsilon$ as follows:
\begin{equation}
\Delta[N] \mid_{\dot{N}=0} = [\int {\cal D} \varepsilon \: \delta(\ddot{\varepsilon})]^{-1} = Det[\frac{d^2}{d \tau^2}].
\end{equation}
Then we arrive at the final form
\begin{equation}
 \Psi_{st} = \int {\cal D}x \, {\cal D}p \, {\cal D}N \,  
 \:[Det \frac{d^2}{d \tau^2}] \: \delta(\dot{N}) e^{iS}. \label{eq:par}
\end{equation}
By using the $\zeta$ function regularization given in appendix A and boundary conditions for $\varepsilon$, we calculate the determinant as
\begin{equation}
Det \frac{d^2}{d \tau^2} =\prod_{n} (\frac{n \pi}{T})^2 = 2T. \label{eq:det}
\end{equation}
By remembering  $N>0$ is used to construct the equation of motion from action (\ref{eq:action2}), 
we present zero-mode integration for $N$.
\begin{equation}
\int {\cal D}N \; \delta(\dot{N}) \: = \: \int^{\infty}_0 dN
\end{equation}

Then we have the path-integration.
\begin{eqnarray}
 \Psi_{st} &=& T \int {\cal D}x \, {\cal D}p \, \int^{\infty}_0 dN   \nonumber \\
& &  \exp \,  [i \int^{T}_{0} d\tau \{ p_i \dot{x}^{i} \, - \, N(p^2 - \omega^2 n^2/c^2) \}] \label{eq:measure}
\end{eqnarray}
 
\section{Wave equation}
Performing the momentum integration, 
we obtain the another form of path-integration expressed only by configuration variables. 
There appears the infinite multiplication of $N$ as the measure. 
But we can absorb such a measure into ${\it D}x$ by changing 
 the time like variable $\tau$ as
\begin{equation}
   d\tau' \equiv  N d\tau,~~ S \equiv  NT.
\end{equation}
This is explicitly explained in appendix B.
Then we obtain
\begin{equation}
 \Psi_{st} = \int^{\infty}_0 dS \,  e^{i\omega^2 S/c^2} \int {\cal D}x \, 
  \exp [i \int^{S}_{0} d\tau' \{ \frac{1}{4} \dot{\bf x}^2 - V(x) \}],\label{eq:schulman}
\end{equation}
where we defined as $ n^2(x) = 1 - V(x) c^2 /\omega^2 $.
Let us rewrite this function as follows:
\begin{equation}
 \Psi_{st} = \int^{\infty}_0 dS \,  e^{i\omega^2 S/c^2} \Psi_0(x=x(S),x_0=x(0),S),
\end{equation}
where $\Psi_0$ is the kernel for non-relativistic Schr\"odinger equation with mass $1/2$.
 \begin{equation}
(i  \frac{d}{dS}+ \nabla^2-V(x)) \Psi_0(x,x_0,S) =\delta^3(x-x_0)\delta(S).
\end{equation}
Then we multiply $ e^{i\omega^2 S/c^2}$ to both hand sides, and integrating by $S$, we have
$$( \frac{\omega^2}{c^2}+\nabla^2-V(x) )\Psi_{st}(x,x_0,\omega) =\delta^3(x-x_0).$$
Since $V(x)=(1-n^2)\omega^2/c^2$, we obtain
 \begin{equation}
( \nabla^2 + n^2\frac{\omega^2}{c^2})\Psi_{st}(x,x_0,\omega) =\delta^3(x-x_0).
\end{equation}
Further, by defining the time dependent function;
 \begin{equation}
\Psi(x,x_0,t)  \equiv \frac{1}{2\pi} \int e^{-i\omega t} \Psi_{st}(x,x_0,\omega) d\omega,
\end{equation}
we have the result
\begin{equation}
( \nabla^2 - \frac{n^2}{c^2}\frac{\partial^2}{\partial t^2})\Psi(x,x_0,t) 
=\delta^3(x-x_0)\delta(t). \label{eq:wave}
\end{equation}

Schulman has done the reverse way. \cite{Schulman} 
He started from wave equation (\ref{eq:wave}), 
and changed to the path integral formulation (\ref{eq:schulman}).
Then he took the stationary point approximation to show the Fermat principle.
This procedure is essentially different from our method.
Our starting point is the Fermat principle itself, 
and by using the technique of quantization procedure of gauge theory, we come to this wave equation.
If the medium has dispersion, the integration by $\omega$ cannot be explicitly done.
The wave equation for dispersive medium is given in integration form.

\section{Conclusion}
We have shown two facts.
First the massless limit of action for relativistic massive particle becomes the observed time along the path of light, 
which realizes the Fermat principle. 
Second the quantization of the system with this action really induces the massless wave equation as light.
The introduction of Fermat Principle from wave equation is well known on many books, 
but in this paper we have shown completely reverse way by using the quantization technique for gauge theory.
The gauge symmetry is the time reparametrization, and  the careful treatment of such symmetry is important to obtain the wave equation.

\begin{appendix}
\subsection{Appendix A}
We show how the zeta-function regularization is performed.
Let the operator $A$, and its eigenvalues as $\lambda_n = a n^x$, where n is the integer and $n>0$.
Then we calculate
\begin{equation}
det A = \prod_{n=1}^{\infty} \lambda_n.
\end{equation}
we find the relation that
\begin{equation}
det A = \exp [tr \{log A \}] = \exp [-\frac{d \zeta(s \mid A)}{ds} \mid_{s=0}],
\end{equation}
where the definition of $\zeta(s \mid A)$ is as follows.
\begin{equation}
\zeta(s \mid A) \equiv \sum_{n=1}^{\infty} \lambda_n^{-s}.
\end{equation}
This can be written as
\begin{equation}
 \zeta(s \mid A)  = \sum_{n=1}^{\infty} a^{-s} n^{-sx}.
 \end{equation}
Therefore,
\begin{equation}
-\frac{d \zeta(s \mid A)}{ds} \mid_{s=0} = \log(a) \zeta(0) - x \zeta'(0),
\end{equation}
where $\zeta(x)$ is the usual zeta function defined as
\begin{equation}
\zeta(x) \equiv \sum_{n=1}^{\infty} n^{-x}.
\end{equation}
We have the formula
\begin{equation}
 \zeta(0) = -1/2, ~~~~~ \zeta'(0) = - (\log 2\pi)/2.
 \end{equation}
Then we obtain
\begin{equation}
-\frac{d \zeta(s \mid A)}{ds} \mid_{s=0} =  \log \frac{(2\pi)^{x/2}}{\sqrt{a}}.
\end{equation}
By using above formula, we obtain
\begin{equation}
det A = \prod_{n=1}^{\infty} a n^x = \frac{(2\pi)^{x/2}}{\sqrt{a}}.
\end{equation}
This is the relation we used in section 3.

\subsection{Appendix B}
After the momentum integration, (\ref{eq:measure}) reduces to
\begin{eqnarray}
 \Psi_{st} &=&T\int^\infty_0 dN \int {\cal D}x \; \; (\prod_{\tau} N^{-d/2}) \nonumber \\
&\times& \exp i\int^{T}_{0} d\tau [\frac{\dot{\vec{x}}^2}{4N}+N(\frac{\omega n}{c})^2],
\end{eqnarray}
where $d$ is the spacial dimension.
Then we remember the explicit form of path-integration by $\vec{x}$, such as
\begin{eqnarray}
\Psi_{st} &=&T \int^\infty_0 dN \lim_{M \to \infty} \int \cdots \int \prod_{i=1}^{d} \prod_{k=1}^{M-1} dx^i(k)(\frac{1}{2\pi i \epsilon N})^{\frac{dM}{2}} \nonumber \\
&\times&\exp i \sum_{k=0}^{M-1}[\frac{(\vec{x}(k+1)-\vec{x}(k))^2}{4N\epsilon} + \epsilon N(\frac{\omega n}{c})^2],
\end{eqnarray}
where $\epsilon \equiv T/M$.  We define 
$$ \bar{\epsilon} = \epsilon N \;\;(d\tau'=N d\tau).$$
Then we obtain
\begin{eqnarray}
\Psi_{st} &=& T \int^\infty_0 dN \lim_{M \to \infty} \int \cdots \int \prod_{i=1}^{d} \prod_{k=1}^{M-1} dx^i(k) \nonumber \\
&\times& (\frac{1}{2\pi i \bar{\epsilon}})^{\frac{dM}{2}} \exp i \sum_{k=0}^{M-1}[\frac{(\vec{x}(k+1)-\vec{x}(k))^2}{4\bar{\epsilon}} + \bar{\epsilon} (\frac{\omega n}{c})^2]\nonumber \\
= &T& \int^\infty_0 dN \int {\cal D}x \; \exp i\int^{NT}_{0} d\tau' [\frac{\dot{\vec{x}}^2}{4}+(\frac{\omega n}{c})^2].
\end{eqnarray}
Let us define $S \equiv NT, \;\; dS \equiv TdN$, then we obtain
$$\Psi_{st} = \int^\infty_0 dS \int {\cal D}x \exp i \int^S_0 d\tau' (\frac{\dot{\vec{x}}^2}{4}+(\frac{\omega n}{c})^2).$$
This is equivalent to equation (\ref{eq:schulman}).

\subsection{Appendix C}
Here we show the BRST construction of path integration (33).
We start with the action (20) with symmetry under (24),(25),(26).
Let us introduce the BRST transformation by the replacement:
\begin{equation}
   \varepsilon(\tau)~~ \to ~~C(\tau),
\end{equation}
where $C$ is the ghost field with odd grassmannian parity.
Then we may have
\begin{equation}
\delta^B x^i = 2 p_i C, ~~ \delta^B p_i =2\frac{\omega^2}{c^2}C n\partial_i n, ~~ \delta^B N = \dot{C}.
\end{equation}
The BRST transformation of ghost field is determined by the nilpotency as
\begin{equation}
\delta^B C =0.
\end{equation}
But the related BRST charge can not be constructed, 
because BRST transformation for $N$ field includes time derivative. 
To solve this difficulty, we must introduce the other fields: 
$\pi$, and $\bar{C}$, and supposing the Poisson brackets
\begin{equation}
\{ N, \pi \}_p = 1, ~~~ \{ C, \bar{C} \}_p = 1.
\end{equation}
Then we write the BRST charge which governs the BRST transformation.
\begin{equation}
Q'_B \equiv C (p^2-(\frac{n\omega}{c})^2) + \pi \dot{C}, ~~~ \delta^B A(\tau) \equiv - \{Q'_B, A(\tau)\}
\end{equation}
It is clear that the form of BRST charge is ambiguous since it includes ``time" derivative of the field. 
To avoid this difficulty we need another kind of ghost field named $F(\tau)$ and
 its conjugate field $\bar{F}$, and require the equation of motion
\begin{equation}
\dot{C} = F.
\end{equation}
Then we have the nilpotent BRST charge
\begin{equation}
Q_B = C (p^2-(\frac{n\omega}{c})^2) + \pi F, ~~~ \{Q_B, Q_B\}_p =0.
\end{equation}
All the variables are
$$x^i, N, C, F ; p_i, \pi, \bar{C}, \bar{F}.$$
Then we have the BRST transformation for each field
\begin{eqnarray}
&&\delta^B x^i = 2 p_i C, ~~ \delta^B p_i =2\frac{\omega^2}{c^2}C n\partial_i n, ~~ \delta^B C = 0, \\
&&\delta^B \bar{C} =-(p^2-(\frac{n\omega}{c})^2), ~~ \delta^B N = F, \\
&&\delta^B F =0, ~~ \delta^B \bar{F} = - \pi, ~~ \delta^B \pi =0.
\end{eqnarray}
It can be seen that the latter 4 variables $(N, \pi, F, \bar{F})$ form the BRST quartet \cite{KUGO}.
Further the action needs to introduce the kinetic term for all the variables 
and also the equation $\dot{C} = F$. The solution is easily found in the form:
\begin{equation}
L = p_i \dot{x}^{i} \, - \, N(p^2 - (\frac{n\omega}{c})^2)\, + \pi \dot{N} 
+ \bar{F} \dot{F} + \bar{C}(\dot{C}- F),
\end{equation}
which is the reasonable form since
\begin{equation}
L = p_i \dot{x}^{i} + \pi \dot{N} + \bar{F} \dot{F} + \bar{C} \dot{C} 
- \{Q_B, N\bar{C}\}.
\end{equation}
Since the BRST transformation is constructed canonically, the  kinetic terms are invariant and the another term
 is also invariant from the nilpotency. 
Therefore the Lagrangian is totally BRST invariant and variation by $\bar{C}$ gives the required equation 
$\dot{C} = F$. We should notice that the gauge is automatically fixed as $\dot{N}=0$.
Now we have the path-integration
\begin{equation}
 \Psi = \int {\cal D}x \, {\cal D}p \, {\cal D}N \, {\cal D}\pi \, {\cal D}C \,  {\cal D}\bar{C}
 \, {\cal D}F \, {\cal D}\bar{F} \, e^{i\int^{T}_{0} L d\tau }.
\end{equation}
The $\bar{C}$ integration requires $\delta(\dot{C}-F)$, and so the $F$ integration changes the term
$$\bar{F} \dot{F} \to \bar{F} \ddot{C}.$$
Then $\bar{F},~ C$ integration gives the determinant $Det \frac{d^2}{d \tau^2}$, and 
$\pi$ integration induces $\delta(\dot{N})$. Therefore we again come to the functional
\begin{eqnarray}
 \Psi & = & \int {\cal D}x \, {\cal D}p \, {\cal D}N \,  
 \:[Det \frac{d^2}{d \tau^2}] \: \delta(\dot{N}) \nonumber \\
&& \exp [i \int^{T}_{0} d\tau \{ p_i \dot{x}^{i} \, - \, N(p^2 -(\frac{n\omega}{c})^2)\}].
\end{eqnarray}
This is the same equation  as (\ref{eq:par}).
So we have the propagator of the massless particle as light.
\end{appendix}

\noindent{\em Acknowledgement.}
\vspace{0.2cm}\\
The author would like to thank Prof. R. Jackiw in MIT for reading through the manuscript and helpful comments.
\\

\end{document}